\documentclass[prl, aps, amsmath, amssymb, twocolumn, superscriptaddress,nofootinbib]{revtex4-1}

\usepackage{graphicx}
\usepackage{hyperref}
\usepackage{physics}
\usepackage{color}
\usepackage{enumerate}
\usepackage{bm,ulem}
 \usepackage[caption=false]{subfig}

\newcommand{\ba}{\begin{eqnarray}}
\newcommand{\ea}{\end{eqnarray}}
\renewcommand{\tr}{\text{Tr}}
\begin{document}

\title{Comment on ``Fluctuations in Extractable Work Bound the Charging Power of Quantum Batteries''}

\author{Stefano Cusumano}
\email[Corresponding author: ]{stefano.cusumano@sns.it}
\affiliation{International Centre for Theory of Quantum Technologies, University of Gda\'nsk, Wita Stwosza 63, 80-308 Gda\'nsk, Poland}

\author{\L ukasz Rudnicki}
\affiliation{International Centre for Theory of Quantum Technologies, University of Gda\'nsk, Wita Stwosza 63, 80-308 Gda\'nsk, Poland}
\affiliation{Center for Theoretical Physics, Polish Academy of Sciences, Aleja Lotnik{\'o}w 32/46, 02-668 Warsaw, Poland}

\begin{abstract}
In the abstract of~[Phys. Rev. Lett. {\bf 125}, 040601 (2020)] one can read  that: [...]{\it to have a nonzero rate of change of the extractable work, the state $\rho_\mathcal{W}$ of the battery cannot be an eigenstate of a "free energy operator", defined by  $\mathcal{F}=H_\mathcal{W}+\beta^{-1}\log \rho_\mathcal{W}$,
where $H_\mathcal{W}$ is the Hamiltonian of the battery and $\beta$ is the inverse temperature} [...].
Contrarily to what is presented below Eq.~(17) of the paper, we observe that the above conclusion does not hold when the battery is subject to nonunitary dynamics.

\end{abstract}

\maketitle
In 
 \cite{chargingpower},
 limits to the charging power of quantum batteries, defined as the rate of variation of the free energy of the battery, are scrutinized. In order to do so, a "free energy operator" is defined, and bounds on the charging power $P(t)=d\langle\mathcal{F}\rangle_\mathcal{W}/dt$, where $\langle\mathcal{F}\rangle_\mathcal{W}=\mathrm{Tr}\left(\mathcal{F}\rho_\mathcal{W}\right)$, are derived. The bounds are used to justify the main conclusion presented in the abstract, namely, bound in Eq. (12) from  \cite{chargingpower} applies to unitary evolution while (16) and (18)  from  \cite{chargingpower} also cover more general Lindblad dynamics. 

We believe the discussed paper presents an interesting analysis of the problem. However, we have found a few mistakes affecting the final conclusions of  \cite{chargingpower}, which we now aim to correct in order to facilitate follow up work which the paper deserves.

In this comment we first rewrite Eq.~(17) of~\cite{chargingpower}, from which the conclusion  under discussion (further called a \textit{hypothesis}) is drawn, to amend for some imprecisions. We then show, using the rewritten equation, that the hypothesis concerning the null charging power of an eigenstate of the free energy operator when the battery is treated as an open system \textit{is not} supported by (16). Then, we perform a direct calculation of the charging power in this special case, to show that the hypothesis indeed does not hold. As a by-product of this analysis, we violate the bound (18) from \cite{chargingpower} showing it is false.


In order to do so we work, as in the original paper, in the eigenbasis of the free energy operator, so that the operators involved in the computation can be written as:
\begin {equation}
\delta\mathcal{F}=\mathcal{F}-\langle\mathcal{F}\rangle_\mathcal{W}=\sum_iw_i\dyad{i},\; \rho_\mathcal{W}=\sum_{i,k}\rho_{ik}\dyad{i}{k},
\label{eq:notation}
\end{equation}
and $L_{j}=\sum_{ik}L_{j}^{ik}\dyad{i}{k}$ are the Lindblad operators.
We are interested in the quantity $\Theta_j:=\langle|[\delta\mathcal{F},L_{j}]|^2\rangle$, where $|A|^2=AA^\dag$. Using Eq.~\eqref{eq:notation} we can explicitly write:
\ba
\comm{\delta\mathcal{F}}{L_{j}}=\sum_{i,k}w_i L_{j}^{ik}\dyad{i}{k}-w_i L_{j}^{ki}\dyad{k}{i},\\
\comm{\delta\mathcal{F}}{L_{j}}^\dag=\sum_{i,k}w_i \left(L_j^{ik}\right)^{*}\dyad{k}{i}-w_i \left(L_j^{ki}\right)^{*}\dyad{i}{k}.
\ea
Some algebra involving matrix multiplication and a careful permutation of indices gives 
 $\left|\comm{\delta\mathcal{F}}{L_{j}}\right|^2=$
\begin{equation}
=\sum_{i,k,\ell}L_j^{ki} \left(L_j^{\ell i}\right)^{*}(w_i^2-w_iw_\ell-w_kw_i+w_\ell w_k)\dyad{k}{\ell},
\end{equation}
from which it is straightforward to get
\begin{equation}
\label{eq:eq17_correct}
\Theta_j=\sum_{i,k,\ell}\rho_{\ell k}L_j^{ki} \left(L_j^{\ell i}\right)^{*}(w_i^2-w_iw_\ell-w_kw_i+w_\ell w_k).
\end{equation}
Eq.~\eqref{eq:eq17_correct} is a slightly corrected variant of Eq.~(17) in~\cite{chargingpower}. 

We are finally ready to verify the case in which the state of the battery is an eigenstate of the free energy operator, i.e., $\rho_\mathcal{W}=\dyad{k_0}$ or $\rho_{\ell k}=\delta_{\ell k_0}\delta_{k_0k}$. Note that the hypothesis under discussion can be proved based on Eq. (16) in~\cite{chargingpower} if and only if $\forall_j \Theta_j=0$. However, we can easily see that if $\rho_{\ell k}=\delta_{\ell k_0}\delta_{k_0k}$ then Eq. (\ref{eq:eq17_correct}) gives
\begin{equation}
\Theta_j=\sum_{i}\left|L_j^{k_0 i}\right|^2(w_i-w_{k_0})^2.
\end{equation}
The right hand side of the above equation is not identically equal to zero, contrarily to what is stated below Eq. (17) in~\cite{chargingpower}. All $\Theta_j$ can simultaneously vanish if and only if $H_\mathcal{W}=w_{k_0} \dyad{k_0}$, or when all $L_j$ act trivially on the range of $H_\mathcal{W}$.

We have just shown that the hypothesis does not follow from Eq. (16) in~\cite{chargingpower}. However, we can also see that, for the master equation Eq.~(14) of~\cite{chargingpower}, the charging power at $t=t_0$ for $\rho_\mathcal{W}(t_0)=\dyad{k_0}$ can explicitly be given
\ba
\label{eq:open_charging_power}
P(t_0)=\sum_j\gamma_j\tr\left(\mathcal{D}_j[\rho_\mathcal{W}(t_0)]H_\mathcal{W}\right),
\ea
where $\mathcal{D}_j[\rho]=L_j\rho L_j^\dag-\acomm{L_j^\dag L_j}{\rho}/2$. This is just because $\mathcal{F}(t_0)=H_\mathcal{W}$, the state $\left|k_0\right\rangle$ is by assumption an eigenstate of $H_\mathcal{W}$, and the time derivative of the von Neumann entropy of $\rho_\mathcal{W}(t)$ at $t=t_0$ is $0$, provided that the dynamics is differentiable \cite{derivative}. We therefore obtain
\ba
\label{eq:open_charging_power}
P(t_0)=\sum_j\gamma_j \sum_{i}\left|L_j^{i k_0}\right|^2(w_i-w_{k_0})\not\equiv0.
\ea
As a matter of fact, $P(t_0)=0$ only when all $\Theta_j$ do vanish. This last conclusion also implies that the bound presented as Eq. (18) in~\cite{chargingpower} is violated.

We  acknowledge support by the Foundation for Polish Science (IRAP project, ICTQT, Contract No. 2018/MAB/5, cofinanced by the EU within the Smart
Growth Operational Programme). We thank  Luis Pedro Garc{\'i}a-Pintos for several clarifications pertaining to the discussed issue.


\begin{thebibliography}{9}
\bibitem{chargingpower}
L. P. Garc{\'i}a-Pintos, A. Hamma, and A. del Campo, \textit{Fluctuations in Extractable Work Bound the Charging Power of Quantum Batteries}, Phys. Rev. Lett. {\bf 125}, 040601 (2020).
\bibitem{derivative} S. Das, S. Khatri, G. Siopsis, and M. M. Wilde, \textit{Fundamental limits on quantum dynamics based on entropy change}, J. Math. Phys. \textbf{59}, 012205 (2017).
\end{thebibliography}
\end{document}